\def\Tr{{\text{Tr}}\,}

\def\be{\begin{equation}}
\def\ee{\end{equation}}
\def\bea{\begin{eqnarray}}
\def\eea{\end{eqnarray}}
\def\bse{\begin{subequations}}
\def\ese{\end{subequations}}

\documentclass[prx,twocolumn,showpacs,amsmath,amssymb,eqsecnum,preprintnumbers]{revtex4}
\usepackage{graphicx} 
\usepackage{dcolumn} 
\usepackage{bm} 
\usepackage{amsfonts}

\draft
\begin{document}
\title{The quantum ferromagnetic transition in a clean Kondo lattice is discontinuous}
\author{T. R. Kirkpatrick$^1$, D. Belitz$^{2,3}$}
\affiliation{$^{1}$ Institute for Physical Science and Technology,
                            University of Maryland, College Park, MD 20742, USA\\
$^{2}$ Department of Physics and Institute of Theoretical Science,
           University of Oregon, Eugene, OR 97403, USA\\
$^{3}$ Materials Science Institute, University of Oregon, Eugene, OR 97403, USA\\
}
\date{\today}

\begin{abstract}
The Kondo-lattice model, which couples a lattice of localized magnetic moments to conduction
electrons, is often used to describe heavy-fermion systems. Because of the interplay between
Kondo physics and magnetic order it displays very complex behavior and is notoriously hard
to solve. The ferromagnetic Kondo-lattice model, with a ferromagnetic coupling between
the local moments, describes a phase transition from a paramagnetic phase to a ferromagnetic
one as a function of either temperature or the ferromagnetic local-moment coupling. At zero
temperature, this is a quantum phase transition that has received considerable attention. It
has been theoretically described 
to be continuous, or second order. Here we show that this belief is
mistaken; in the absence of quenched disorder the quantum phase transition is first order,
in agreement with experiments,
as is the corresponding transition in other metallic ferromagnets.
\end{abstract}
\maketitle

\section{Introduction}
\label{sec:1}

The Kondo-lattice model, which is the standard model for heavy-fermion systems, consists of a lattice of interacting 
localized spins or local moments coupled to conduction electrons \cite{Doniach_1977}. In a typical heavy-fermion
material, the local moments are due to f-electrons, and the conduction electrons populate one or more separate
bands. At least two physically distinct quantum phase transitions can occur in such a system as a function of the
relative strength of the Kondo coupling between the local moments and the conductions electrons on one hand,
and the intra-local-moment coupling on the other. One is a transition from a magnetically ordered phase to a 
nonmagnetic one; the other, a transition from an ordinary Fermi liquid with a small Fermi surface that is comprised 
of the conduction electrons only to a heavy Fermi liquid with a large Fermi surface. The latter transition is produced 
by a hybridization of the f-electrons with the conduction electrons \cite{Gegenwart_Si_Steglich_2008}. 
These two transitions are generically expected to be separate, but they may coincide in some materials. 
In most heavy-fermion materials the
coupling between the local moments is antiferromagnetic \cite{Gegenwart_Si_Steglich_2008}, but over
the years an increasing number of heavy-fermion metals have been discovered that display ferromagnetic
order, see Refs.~\cite{Yamamoto_Si_2010, Brando_et_al_2015} and references therein. This raises the issue of a ferromagnetic
quantum phase transition in these complex metallic systems.

Two obvious questions regarding this ferromagnetic transition are: (1) What are the properties of this transition?
In particular, is it continuous (second order) or discontinuous (first order)? (2) What happens to the Fermi surface
of the conduction electrons across the transition? If the hybridization transition that delocalizes the f-electrons
coincides with the ferromagnetic transition, then the latter will be accompanied by a concurrent transition from
a small Fermi surface to a large one, or vice versa \cite{Fermi_surface_footnote}; 
%
otherwise, there will be a succession of magnetic and
hybridization transitions. In this paper we answer the first question: The transition
is first order. We will comment on the second question in the discussion, Sec.~\ref{sec:3}.

The ferromagnetic quantum phase transition in metals in general has an even longer history than the Kondo effect, going back
to Stoner's mean-field theory \cite{Stoner_1938}. Hertz \cite{Hertz_1976} used itinerant ferromagnets as
an example for his more general theory of quantum phase transitions. He concluded that the ferromagnetic quantum
phase transition in clean metals is continuous and mean-field like in all dimensions $d>1$. It later became clear
that this conclusion is not correct \cite{Belitz_Kirkpatrick_Vojta_1999, Kirkpatrick_Belitz_2012b}. The problem
lies in soft or gapless particle-hole excitations in the conduction-electron system that couple to the magnetic
fluctuations. Hertz theory takes this coupling into account in an approximation that correctly describes Landau 
damping, but is not sufficient for correctly describing the nature of the magnetic quantum phase transition.
In Ref.~\cite{Belitz_Kirkpatrick_Vojta_1999} it was shown that the zero-temperature ($T=0$) transition in a clean ferromagnet
in dimensions $d=2$ and $d=3$ is {\em generically} first order \cite{generic_footnote}. 
The physical mechanism behind
this conclusion is in many ways analogous to the well-known fluctuation-induced first-order transition in
superconductors and liquid crystals \cite{Halperin_Lubensky_Ma_1974}, with the soft fermionic excitations
playing the role of the photons (in the case of superconductors) or the nematic Goldstone mode (in the case
of liquid crystals). An important difference, however, is that the (fictitious) quantum ferromagnetic transition in the 
absence of any coupling to the conduction electrons, which is described by Hertz theory, is above its upper critical
dimension $d_c^+=1$, as opposed to the transition in a superconductor or liquid crystal, which is below its
upper critical dimension $d_c^+=4$. As a result, the fluctuation-induced first-order transition in quantum
ferromagnets is expected to be much more robust again order-parameter fluctuations than its classical
counterparts \cite{Brando_et_al_2015}. At $T>0$ the fermionic soft modes acquire a mass, the 
fluctuation-induced first-order transition becomes weaker, and with increasing temperature a tricritical
point appears in the phase diagram above which the transition is continuous and belongs to the applicable
classical universality class \cite{Belitz_Kirkpatrick_Rollbuehler_2005}.

The initial theory of the first-order quantum phase transition \cite{Belitz_Kirkpatrick_Vojta_1999} was for 
itinerant ferromagnets. It was later shown that its conclusions hold much more generally and apply to
clean metallic ferromagnets irrespective of whether the ferromagnetism is due to the conduction electrons,
or due to electrons in a separate band, and even to ferrimagnets and magnetic nematics
\cite{Kirkpatrick_Belitz_2012b, Kirkpatrick_Belitz_2011a}. This generalization of the theory is
important, since many systems in which a discontinuous ferromagnetic quantum phase transition
is observed are not itinerant ferromagnets \cite{Brando_et_al_2015}. It also raises the question
whether Kondo lattices, with their complicated interplay between Kondo physics and magnetic order,
are different in this respect from other metallic quantum ferromagnets. Mean-field approaches
\cite{Irkhin_Katsnelson_1990, Perkins_et_al_2007, Irkhin_2014} generically (i.e., for a simple
band structure) yield a second-order transition, although special features in the density of states
can lead to a first-order transition \cite{Irkhin_Katsnelson_1990, Irkhin_2014}.
A second-order transition also is implicit in the renormalization-group (RG) analysis of 
Ref.~\cite{Yamamoto_Si_2010}. Here we show that these analyses gave the wrong answer for the order
of the transition for the same reason as Hertz theory, and that a careful consideration of 
conduction-electron fluctuations
leads to a first-order transition, as it does in all other metallic quantum
ferromagnets.

\section{Theory of the ferromagnetic quantum phase transition in a Kondo lattice}
\label{sec:2}

\subsection{The model}
\label{subsec:2.1}

In the general Kondo-lattice problem the local moments are coupled to each other by an exchange coupling $I$ that can be 
either antiferromagnetic or ferromagnetic. We are interested in the latter, so we take $I<0$. Similarly, the coupling $J$ between 
the local moments and the conduction electrons can have either sign. As we will see, the order of the ferromagnetic transition 
is independent of the sign of $J$, although the physics related to Kondo screening and the hybridization of f-electrons and
conduction electrons crucially depend on it. For our purposes 
we thus do not specify the sign of $J$ for now and will come back to this issue in the discussion.

We start with a standard Hamiltonian description of the Kondo-lattice problem. The Hamiltonian consists of three parts,
\bse
\label{eqs:1}
\be
{\hat H} = {\hat H}_{\text{FL}} + {\hat H}_{\text{LM}} + {\hat H}_{\text{c}}
\label{eq:1a}
\ee
Here ${\hat H}_{\text{FL}}$ is a Fermi-liquid Hamiltonian that describes the conduction electrons. For simplicity, we consider
only one conduction-electron band,
\be
{\hat H}_{\text{FL}} = \sum_{\bm{k},\sigma} \left(\epsilon_{\bm{k}} - \mu\right) {\hat c}^{\dagger}_{\bm{k},\sigma} {\hat c}_{\bm{k},\sigma}
                 + {\hat H}_{\text{int}}
\label{eq:1b}
\ee
where the ${\hat c}^{\dagger}_{\bm k}$ and ${\hat c}_{\bm k}$ are fermionic creation and annihilation operators, 
respectively, and $\epsilon_{\bm{k}}$
is the single-electron energy-momentum relation. Here we describe the conduction electrons by Bloch states
with a wave number ${\bm k}$. $\sigma$ is the spin index, $\mu$ is the chemical potential, and ${\hat H}_{\text{int}}$
describes the electron-electron interaction. The latter is important for what follows, as we will discuss below. 
${\hat H}_{\text{LM}}$ describes local moments ${\bm S}$ on real-space sites $i,j$ that interact via a ferromagnetic nearest-neighbor 
interaction $I<0$:
\be
{\hat H}_{\text{LM}} = I \sum_{\langle i,j\rangle} {\bm S}_i\cdot{\bm S}_j\ .
\label{eq:1c}
\ee
Finally, ${\hat H}_c$ describes the coupling between the local moments and the
spin density of the conduction electrons with coupling constant $J$,
\be
{\hat H}_{\text c} = J\sum_i {\bm S}_i\cdot {\hat c}^{\dagger}_{i,\sigma}{\bm\sigma}_{\sigma,\sigma'} {\hat c}_{i,\sigma'}\ .
\label{eq:1d}
\ee
\ese
Here $\bm\sigma$ denotes the Pauli matrices, and the ${\hat c}_i^{\dagger}$ and ${\hat c}_i$ are creation and
annihilation operators, respectively, for conduction electrons at site $i$.

\subsection{Effective field theory}
\label{subsec:2.2}

In order to study the phase transition in the local-moment subsystem we are interested in, we
next rewrite the partition function
\bse
\label{eqs:2}
\be
Z = \Tr e^{-{\hat H}/T}
\label{eq:2a}
\ee
in terms of a functional integral
\be
Z = \int D[{\bar\psi},\psi]\,D[\bm M]\,e^{S_{\text{FL}}[{\bar\psi},\psi] + S_{\text{LM}}[{\bm M}] + S_{\text{c}}[{\bar\psi},\psi;\bm M]}\ .
\label{eq:2b}
\ee
\ese
Here $S$ is the action of an effective field theory whose three parts correspond to the three parts of the Hamiltonian.
We now specify and discuss them one by one.

The local-moment part takes the form of a quantum $\phi^4$-theory \cite{sigma_model_footnote}
%
%
\bea
S_{\text{LM}}[{\bm M}] &=& -\int dx\,\left[t\,{\bm M}^2(x) + (\nabla{\bm M})^2(x) + u\,({\bm M}^2(x))^2 \right] 
\nonumber\\
&& + S_{\text{LM}}^{\text{dyn}}[{\bm M}]
\label{eq:3}
\eea
Here ${\bm M}$ is the order-parameter (OP) field, $x = ({\bm x},\tau)$ comprises the real-space and imaginary-time
coordinates, and $\int dx = \int d{\bm x} \int_0^{1/T} d\tau$. The term $S_{\text{LM}}^{\text{dyn}}[{\bm M}]$
describes the bare dynamics of the local moments. Deep inside the ordered phase it takes the form of a
Wess-Zumino or Berry-phase term (see, e.g., Ref.~\cite{Altland_Simons_2010})
\be
S_{\text{LM}}^{\text{dyn,WZ}}[{\bm M}] = -i\langle\vert{\bm M}\vert\rangle \int dx {\bm A}[{\bm M}]\,\partial_{\tau}{\bm M}(x)\ .
\label{eq:4}
\ee
To lowest order in the order parameter, and for a magnetization pointing in the $z$-direction, the vector potential ${\bm A}$ has
the form ${\bm A}[{\bm M}] \approx (-M_y, M_x,0)$. Physically, such a term describes the Bloch precession of the local moments,
and therefore it must also be present in the soft-spin or LGW formulation of the action $S_{\text{LM}}$ given above. However,
the coupling to the fermions produces other dynamical terms, the most important of which is the Landau-damping term which,
in Fourier space, takes the form \cite{Hertz_1976}
\be
S_{\text{LM}}^{\text{dyn,L}}[{\bm M}] = -\sum_{{\bm k},\omega} {\bm M}({\bm k},\omega) (\vert\omega\vert/\vert{\bm k}\vert)
    \cdot{\bm M}(-{\bm k},-\omega)\ .
\label{eq:5}
\ee
This term dominates the Berry-phase term, as well as other dynamical terms generated by the coupling between the order 
parameter and the fermions. This is true both in the paramagnetic phase and at any putative quantum critical point,
irrespective of the nature of the quantum phase transition, as is obvious from power counting. As we will
show below, the ferromagnetic quantum phase transition is actually first order, which can be established without 
considering any dynamical term in $S_{\text{LM}}$.

The fermionic sector is described by a standard action $S_{\text{FL}}$ for a Fermi liquid. ${\bar\psi}$ and $\psi$ are 
fermionic spinor fields, and
$S_{\text{FL}}$ consists of a term bilinear in $\bar\psi$ and $\psi$ that describes band electrons with electron-momentum
relation $\epsilon_{\bm k}$, and four-fermion terms that describe the electron-electron interaction. The latter contains in
particular a spin-triplet interaction of the form
\bse
\label{eqs:6}
\be
S_{\text{int}}^t = \Gamma_t \int dx\ {\bm n}_s(x) \cdot {\bm n}_s(x)\ .
\label{eq:6a}
\ee
Here 
\be
{\bm n}_s(x) = {\bar\psi}(x){\bm\sigma}\psi(x)
\label{eq:6b}
\ee
\ese
is the electronic spin density, and 
$\Gamma_t$ is the spin-triplet interaction amplitude, which for simplicity we consider static and
point-like. Since we are not interested in systems where the conduction electrons
by themselves develop magnetic order, we assume that $\Gamma_t$ is small enough for the system to not
be an itinerant ferromagnet.

If one aims to construct a complete effective field theory one can express the fermionic degrees of freedom
in terms of bosonic ones that are isomorphic to bilinear products of $\bar\psi$ and $\psi$, and that capture
the soft modes in the fermion sector \cite{Belitz_Kirkpatrick_2012}. This would be the preferred strategy if
one wanted to perform a systematic RG study of the effective field theory, since the
renormalization of the fermionic sector is rather involved if it is formulated in terms of fermionic fields.
However, it turns out that, remarkably, one can establish the first-order transition of the ferromagnetic quantum phase
transition by using established properties of Fermi liquids without an explicit formulation of the fermionic sector.
(Deriving these properties in the first place does, of course, require an explicit formulation.) We therefore do not 
dwell on the detailed form of the fermionic sector of the effective field theory.

Finally, the coupling between the local moments and the fermions is described by Eq.~(\ref{eq:1d}). In the language 
of the effective field theory, this takes the form
\be
S_{\text{c}}[{\bar\psi},\psi;\bm M] = -J \int dx\ {\bm M}(x)\cdot {\bm n}_s(x)\ .
\label{eq:7}
\ee

\subsection{Order-parameter action}
\label{subsec:2.3}

We now define an effective order-parameter action by formally integrating out the fermions. The partition
function then takes the form
\bse
\label{eqs:8}
\be
Z = \int D[{\bm M}]\ e^{S_{\text{eff}}[{\bm M}]}\ ,
\label{eq:8a}
\ee
where we have defined
\bea
S_{\text{eff}}[{\bm M}] &=& S_{\text{LM}}[{\bm M}] + \ln \int D[{\bar\psi},\psi]\ e^{S_{\text{FL}}[{\bar\psi},\psi] 
                                            + S_{\text c}[{\bar\psi},\psi;{\bm M}]}
\nonumber\\
                                   &=& S_{\text{LM}}[{\bm M}] + \delta S[{\bm M}]\ .                                           
\label{eq:8b}
\eea
\ese
In the second line in Eq.~(\ref{eq:8b}), $\delta S$ describes the effects of the fermions on the local moments.
Notice that this is purely formal, as the fermionic integral cannot be performed unless the fermions are noninteracting.
However, as we will see, Eq.~(\ref{eq:8b}) is very useful for utilizing known properties of Fermi liquids for obtaining
information about the local moments.

\subsection{Free energy, and first-order transition}
\label{subsec:2.4}

We now show that fluctuations in the fermion sector cause the ferromagnetic quantum phase transition
in the local-moment system to be first order or discontinuous.

\subsubsection{Renormalized Landau theory}
\label{subsubsec:2.4.1}

In order to determine whether a phase transition is continuous or discontinuous, one needs to 
consider the free energy. In the simplest approximation this can be done by treating the order
parameter in a mean-field approximation. In the current context, this amounts to replacing the
fluctuating magnetization ${\bm M}(x)$ by an $x$-independent magnetization $M$ that
we take to point in the 3-direction. We will discuss the validity of this approximation at the end of
Sec. \ref{sec:2}. Denoting the 3-component of ${\bm n}_{\text{s}}$ by 
$n_{\text{s}}$, the second term in Eq.\ (\ref{eq:8b}), which describes the effect of the 
coupling between the fermions and the OP, can be written
\bea
\delta S[M] &=& \ln \int D[{\bar\psi},\psi]\ e^{S_{\text{FL}}[{\bar\psi},\psi] - J M \int dx\,n_{\text{s}}(x)}
\nonumber\\
        &=& \ln \left\langle e^{-J M \int dx\,n_{\text{s}}(x)} \right\rangle_{\text{FL}}\ ,
\label{eq:9}
\eea
where in the second line we have dropped a constant contribution to the action, and 
$\langle\ldots\rangle_{\text{FL}}$ denotes an average with respect to the action $S_{\text{FL}}$.

Now consider the longitudinal spin susceptibility $\chi(h)$ of fermions described by the action $S_{\text{FL}}$ and
subject to a magnetic field $h$. It is given by a two-point spin-density correlation function:
\be
\chi(h) = \frac{T}{V}\int dx\,dy\,\langle\delta n_{\text{s}}(x)\,\delta n_{\text{s}}(y)\rangle_{S_{\text{h}}}\ ,
\label{eq:10}
\ee
where $S_{\text{h}} = S_{\text{FL}} + h\int dx\,n_{\text{s}}(x)$, and 
$\delta n_{\text{s}}(x) = n_{\text{s}}(x) - \langle n_{\text{s}}(x) \rangle_{S_{\text{h}}}$.
By differentiating Eq.\ (\ref{eq:9}) twice with respect to $M$ we find
\be
\frac{d^2\,\delta S}{dM^2} = \frac{V}{T}\,J^2 \chi(J M)\ .
\label{eq:11}
\ee
Dropping an irrelevant constant contribution to $\delta S$, we have $\delta S[M=0] = 0$
Furthermore, since the fermion sector is not magnetically ordered, we also have $d\,\delta S/dM\vert_{M=0} = 0$.
Integrating Eq.~(\ref{eq:11}) thus yields
\be
\delta S[M] = \frac{V}{T}\,J^2 \int_0^M dm_1 \int_0^{m_1} dm_2\ \chi(J m_2)\ .
\label{eq:12}
\ee
$S_{\text{LM}}$ has the usual Landau form of a power series in powers of $M^2$, and all
dynamical terms vanish. The complete renormalized Landau free-energy density $f_{\text{eff}} = -(T/V)S_{\text{eff}}$
thus is
\bse
\label{eqs:13}
\be
f_{\text{eff}}[M] = t\,M^2 + \delta f[M] + u\,M^4 + O(M^6)\ .
\label{eq:13a}
\ee
Here $t$ and $u$ are Landau parameters, and
\be
\delta f[M] = -J^2 \int_0^M dm_1 \int_0^{m_1} dm_2\ \chi(J m_2)\ .
\label{eq:13b}
\ee
\ese

This result expresses the correction to the usual Landau action in terms of the spin susceptibility
of nonmagnetic fermions in the presence of an effective homogeneous magnetic field given by $J M$.
It is a ``renormalized Landau theory'' in the sense that it includes the effects of fluctuations extraneous
to the order-parameter fluctuations. The remaining question is the behavior of the susceptibility $\chi$ that 
represents these fluctuations for small $M$. The salient point is that 
$\chi$ is not an analytic function
of $M$ at $M=0$.

\subsubsection{Effective free energy}
\label{subsubsec:2.4.2}

It is well known that various observables in a Fermi liquid are
nonanalytic functions of the temperature. For instance, the specific heat coefficient
has a $T^2\ln T$ term \cite{Baym_Pethick_1991}. The spin susceptibility in
a three-dimensional system was found to have no such nonanalytic behavior
\cite{Carneiro_Pethick_1977}. However, this absence of a nonanalyticity was later shown to 
be accidental, and to pertain only to the $T$-dependence in three dimensions. 
In dimensions $d\neq 3$ there is a $T^{d-1}$ nonanalyticity, and even in $d=3$ at $T=0$ the inhomogeneous spin susceptibility has
a $k^2\ln k$ wave-number dependence \cite{Belitz_Kirkpatrick_Vojta_1997, Chitov_Millis_2001b, 
Chubukov_Maslov_2003}. This nonanalyticity is a consequence of soft modes
that exist at zero temperature in any Fermi liquid. From general 
scaling arguments one expects a corresponding nonanalyticity for the
homogeneous susceptibility at $T=0$ 
as a function of a magnetic field $h$, namely,
$\chi(h) \propto \text{const.} + h^{d-1}$ in generic dimensions, and 
$\chi(h) \propto \text{const.} - h^2 \ln h$ in $d=3$. These scaling arguments have
been shown to be exact, as far as the exponent is concerned, by a RG 
treatment \cite{Belitz_Kirkpatrick_2014}, and they are consistent with explicit perturbative
calculations \cite{Misawa_1971, Barnea_Edwards_1977, Betouras_Efremov_Chubukov_2005}. The sign of the effect is universal
and can be established as follows. Fluctuations suppress the tendency of a
Fermi liquid to order ferromagnetically, and therefore the fluctuation correction to the
bare zero-field susceptibility is negative, $\delta\chi(0) < 0$. A 
magnetic field
suppresses the fluctuations, and therefore $\delta\chi(h) - \delta\chi(0) > 0$. This implies
that the nonanalyticity in $\chi(h\to 0)$ has a positive sign:
\be
\chi(h\to 0) = \chi(0) + \begin{cases} a_d\, h^{d-1} & \text{for $1<d<3$} \\
                                                          a_3\,h^2\ln(1/h) & \text{for $d=3$}
                                   \end{cases}\ ,
\label{eq:14}
\ee
where $a_d>0$. For the renormalized Landau free-energy density, Eq.\ (\ref{eq:13a}), we
thus obtain
\bea
f_{\text{eff}}[M]  &=& -h\,M + t\,M^2 + u\,M^4 
\nonumber\\
&&\hskip -0pt - v_d\times\begin{cases} M^{d+1} + u\,M^4 & (1<d<3)\\
                                                  M^4\ln(1/M) & (d=3)\ .
                             \end{cases} \hskip 20pt
\label{eq:15}
\eea
Here $v_d \propto \vert J\vert^{d+1} > 0$, and we have added an external magnetic field $h$. For $d=3$ this result was first derived
in Ref.~\cite{Belitz_Kirkpatrick_Vojta_1999} in the context of itinerant ferromagnets. The current derivation shows that it
is completely general and applies to all metallic quantum ferromagnets, including Kondo lattices. The negative term in the 
free energy, which dominates the quartic term for all $d\leq 3$, necessarily leads to a first-order
ferromagnetic transition. We stress that while this is a fluctuation-induced first-order
quantum phase transition, the relevant fluctuations are {\em not} the 
OP fluctuations, but are fermionic in nature. For purposes of an analogy with the well-known
classical fluctuation-induced first-order transitions \cite{Halperin_Lubensky_Ma_1974},
the latter play a role that is analogous to that of the vector potential in superconductors,
or the director fluctuations at the nematic-smectic-A transition. An important difference,
which was already mentioned in the Introduction, is that in these classical systems the OP 
fluctuations are below
their upper critical dimension, which makes them strong enough to make the first-order
transition 
weak and hard to observe at best, and destroy it altogether at worst \cite{Anisimov_et_al_1990}.
By contrast, in the case of a quantum FM 
the OP 
fluctuations are
{\em above} their upper critical dimension, so the first-order transition predicted
by the renormalized Landau theory will 
be much more robust.

\begin{figure}[t]
\begin{center}
\includegraphics[width=0.8\columnwidth,angle=0]{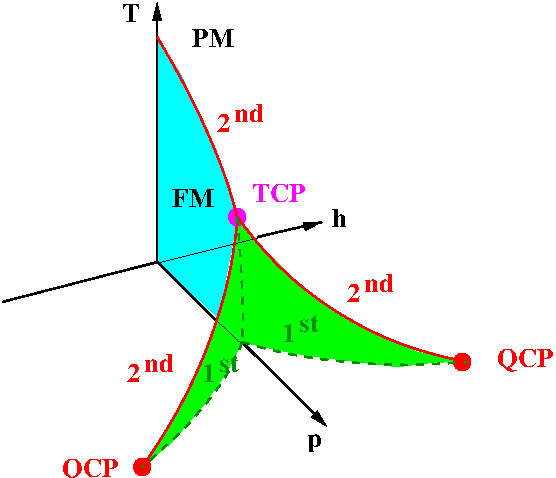}
\end{center}
\caption{Schematic phase diagram in the space spanned by temperature ($T$),
              hydrostatic pressure ($p$), and magnetic field ($h$). Shown are the 
              FM and 
              PM phases, the tricritical point (TCP), and the two quantum critical points (QCP). 
              Solid and dashed lines denote second- 
              and first-order transitions,
              respectively. The tricritical wings emerging from the TCP are surfaces of first-order
              transitions. }
\label{fig:1}
\end{figure}

A nonzero temperature cuts off the magnetic-field singularity \cite{Betouras_Efremov_Chubukov_2005},
and with increasing temperature the fluctuation-induced term in the free energy becomes less and
less negative. Suppose the Landau parameter $t$ at $T=0$ is a monotonically increasing function of, 
say, hydrostatic pressure $p$, and let $t(p=0,T=0) < 0$. Then there will be a QPT 
at some nonzero pressure $p_{\text{c}}$. As the transition temperature
is increased from zero by lowering $p$, one 
expects a tricritical point in the phase diagram. 
Below the tricritical temperature the transition will be discontinuous due to the mechanism described 
above, while at higher temperatures it will be continuous. In the presence of an external magnetic
field there appear surfaces of first-order transitions, or tricritical wings \cite{Belitz_Kirkpatrick_Rollbuehler_2005}, 
as is the case for any phase diagram that contains a tricritical point \cite{Griffiths_1970, Griffiths_1973}.
The phase diagram has the schematic structure shown in the right-most panel in Fig.~\ref{fig:1}.%
%
%

\section{Discussion}
\label{sec:3}

The mechanism we have described that leads to a first-order quantum ferromagnetic transition is very general:
It is a universal long-wavelength effect that depends only on the presence of conduction electrons that form a
Fermi liquid. It thus is valid for any metallic ferromagnet with a Fermi surface. A comparison with other
theoretical analyses that have come to different conclusions for the special case of a Kondo lattice is therefore called for, and
we start our discussion with this.

Mean-field approximations 
on Eqs.~(\ref{eqs:1}) have been used to develop a Stoner-like theory of ferromagnetism on a Kondo 
lattice \cite{Irkhin_Katsnelson_1990, Perkins_et_al_2007, Irkhin_2014}. As one would 
expect within such an approach, it generically yields a  transition that is continuous with mean-field critical exponents,
although the transition may be first order due to special features in the density of states \cite{Irkhin_Katsnelson_1990, Irkhin_2014}.
Later, Yamamoto and Si \cite{Yamamoto_Si_2010} used a combination of field theoretic and RG
methods in an attempt to understand the fate of the Kondo screening in the ferromagnetic phase. For an antiferromagnetic
Kondo coupling $J>0$, and for $J \ll \vert I \vert$, they concluded that the Kondo coupling flows to zero, the Kondo screening in
the ferromagnetic phase breaks down, and the system has a small Fermi surface. While their focus was on the stable fixed point
that describes the ferromagnetic phase,
their results imply that the ferromagnetic quantum phase transition is second order. That is, the RG treatment does
not change the order of the phase transition compared to the generalized Stoner theory of Ref.~\cite{Perkins_et_al_2007}.
Two related points are: (1) The analysis of Ref.~\cite{Yamamoto_Si_2010} does not yield the nonanalytic wave-number dependence
found in Ref.~\cite{Chubukov_Pepin_Rech_2004} for a closely related model. (2) It finds a linear magnetization dependence
for the spin-wave stiffness coefficient $D(m)$ in the magnon dispersion relation $\Omega = D(m) {\bm k}^2$, whereas 
Ref.~\cite{Belitz_et_al_1998} found a nonanalytic $m$-dependence of $D$ for a quantum nonlinear sigma model. All of these
effects, as well as the nonanalytic field dependence of the spin susceptibility of a Fermi liquid, Eq.~(\ref{eq:14}), have the same origin, 
and lead to a first-order quantum ferromagnetic transition as discussed in Sec.~\ref{sec:2}. These discrepancies can be traced back
to the fact that the starting action of Ref.~\cite{Yamamoto_Si_2010} does not contain any interactions for the conduction
electrons, and the latter are responsible for the nonanalyticities that in turn lead to the first-order transition. In principle,
a complete RG analysis will generate an electronic interaction, via an exchange of excitations in the local-moment system to
which the conduction electrons are coupled, even if none was included in the bare action. However, this requires the consideration
of terms of higher-loop order than were kept in Ref.~\cite{Yamamoto_Si_2010}. The fluctuations that cause the first-order
transition discussed in Sec.~\ref{sec:2} were thus effectively not considered. 

We conclude with a number of additional remarks:
\begin{enumerate}
\item Our conclusion that the ferromagnetic quantum phase transition in clean heavy-fermion or Kondo-lattice systems is first order 
         is in agreement with experimental results. Examples include UGe$_2$ \cite{Huxley_et_al_2001, Huxley_et_al_2007}, 
         URhGe \cite{Aoki_et_al_2011b, Huxley_et_al_2007}, and UCoGe \cite{Aoki_et_al_2011b, Hattori_et_al_2010}.
\item As already mentioned in Sec.~\ref{subsec:2.2}, there are a number of dynamic processes in the Kondo-lattice problem, 
   and corresponding frqeuency-dependent terms in the action. One is the Berry-phase term that describes the Bloch precession
   of the local moments, Eq.~(\ref{eq:4}). It is physically obvious, although not commonly appreciated, that such a term must 
   also exist in the action for an itinerant ferromagnet, in which case it is generated by the dynamics of the conduction electrons.
   In addition, there is a term describing relaxation processes that do not become slow in the limit of small wave numbers. 
   At a continuous quantum phase transition, all three of these terms are irrelevant compared to the Landau-damping term
   given by Eq.~(\ref{eq:5}).
\item It follows from thermodynamic considerations alone, namely, from various Clapeyron-Clausius relations, that the tricritical wings
   shown in Fig.~\ref{fig:1} are perpendicular to the $T=0$ plane, but not perpendicular to the $p$-axis and point in the direction of the 
   paramagnetic phase \cite{Kirkpatrick_Belitz_2015a}. These features, as well as the overall structure of the phase diagram, are
   in excellent agreement with experimental observations \cite{Brando_et_al_2015}.
\begin{figure*}[t]
\centerline{\includegraphics*[width=1.8\columnwidth]{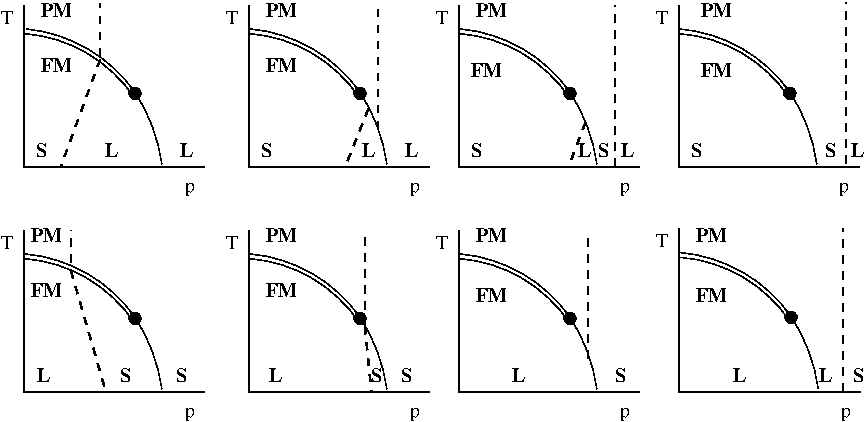}}
\caption{Possible temperature - pressure ($T$-$p$) phase diagrams for systems with a small (first row) or large (second row) 
              Fermi surface deep inside the ferromagnetic (FM) phase. Double and single solids lines represent 
              continuous and first-order transitions, respectively, from a paramagnetic (PM) phase to a FM one,
              separated by a tricritical point (dot). Dashed lines 
              represent the hybridization temperature whose pressure dependence has been neglected for simplicity. 
              S and L indicate regions of small and large Fermi surfaces, respectively.}
\label{fig:2}
\end{figure*}
\item Quenched non-magnetic disorder changes the nature of the fermionic soft modes from ballistic to diffusive. More importantly,
   it changes the sign of the non-analytic term in Eqs.~(\ref{eq:14}) and (\ref{eq:15}). This is because the interplay of electron-electron
   interactions and quenched disorder enhances the electronic spin susceptibility
   compared to the Pauli susceptibility \cite{Altshuler_Aronov_1984, Finkelstein_1983}.
   That is, the combined disorder and interaction fluctuations increase the zero-field susceptibility. A magnetic field again suppresses
   the fluctuations, and hence the sign of the nonanalyticity in $\chi_h\to 0)$ changes compared to Eq.~(\ref{eq:14}):
   \be
   \chi(h\to 0) = \chi(0) - {\tilde a}_d\,h^{(d-2)/2}\qquad (2<d<4)\ ,
   \label{eq:16}
   \ee
   with ${\tilde a}_d > 0$. The changed exponent reflects the diffusive nature of the soft modes. As a result, the
   transition within the renormalized mean-field theory is second order with non-mean-field exponents. The evolution
   of the phase diagram with increasing disorder strength has been discussed in Ref.~\cite{Sang_Belitz_Kirkpatrick_2014},
   and the asymptotic and pre-asymptotic critical behavior in Ref.~ \cite{Kirkpatrick_Belitz_2014}. 
%
%
\item For 
   an antiferromagnetic Kondo coupling $J$, the existence or otherwise of Kondo screening in the 
   FM phase needs to be reconsidered. The scale dimension of the Kondo coupling, which determines the answer, 
   depends on the dynamical scale dimension of the order-parameter field \cite{Yamamoto_Si_2010}, which in turn depends 
   on the same physics that determines the order of the phase transition. The RG analysis of Ref.~\cite{Yamamoto_Si_2010}
   concluded that deep inside the FM phase the Kondo screening generically breaks down, i.e., the Fermi surface
   is small. For FM systems that are driven paramagnetic by the application of pressure, such as UGe$_2$, this
   would imply that there is a hybridization transition from a small Fermi surface to a large one with increasing pressure.
   Assuming that an applied field, or a spontaneous magnetization, favors a large Fermi surface at least in a range
   of fields, as is the case in the antiferromagnetic heavy-fermion metal CeRhIn$_5$ \cite{Jiao_et_al_2015} (see also the
   phenomenological theory of Ref.~\cite{Yang_Pines_2014}), this suggests possible phase diagrams as shown in the first
   row of Fig.~\ref{fig:2}. However, for antiferromagnetic Kondo lattices it is known that hydrostatic pressure can either
   favor or suppress hybridization in different materials \cite{Yang_Pines_2014}, and for UGe$_2$ there is some experimental
   evidence for a large Fermi surface deep inside the FM phase and a small one close to the first-order
   transition \cite{Huxley_et_al_2003}. Possible phase diagrams for this case are shown in the second row of Fig.~\ref{fig:2}.
   If a field or magnetization favors a small Fermi surface, as must be the case at least in very strong fields and is observed
   in, e.g., the antiferromagnet CeRu$_2$Si$_2$ \cite{van_der_Meulen_et_al_1991} (see also, e.g.,
   Refs.~\cite{Belitskii_Goltsev_1989, Kusminskyi_et_al_2008}), the position and direction of the dashed line
   within the FM phase in Fig.~\ref{fig:2} will change in obvious ways.
   We also note that the hybridization transition is a true phase transition only at $T=0$, and a crossover at $T>0$. However,
   even a crossover could trigger a metamagnetic transition with a sharp discontinuity in the magnetization. This raises
   the possibility that the FM2-to-FM1 metamagnetic transition observed in UGe$_2$ is a signature of the hybridization
   transition \cite{Huxley_et_al_2003}. The critical behavior at the critical point where the line of first-order metamagnetic
   transitions ends has been analyzed in Ref.~\cite{Millis_et_al_2002}.
\item If the Fermi surface deep inside the FM phase is large (small) and if a magnetization enhances (suppresses) hybridization, 
  then there will be a whole region of parameter values for
  which the magnetic transition and the hybridization transition coincide, as can be seen from the third panel in the
  second row of Fig.~\ref{fig:2}. In the case of a first-order magnetic transition the two transitions can thus coincide
  generically, whereas for a second-order transition this can happen only for a set of parameter values that is of
  measure zero.

\end{enumerate}

\begin{acknowledgements}
This work was supported by the NSF under Grants No. DMR-1401410 and No. DMR-1401449. 
\end{acknowledgements}


\begin{thebibliography}{49}
\bibitem[{Fer()}]{Fermi_surface_footnote}
\bibinfo{note}{Ferromagnetic order splits the Fermi surface; therefore,
  ``small'' or ``large'' in this context refers to the combined Fermi
  surfaces.}
\bibitem[{gen()}]{generic_footnote}
\bibinfo{note}{The first-order nature of the transition is due to the
  fundamental soft-mode structure of metals, and hence independent of the band
  structure and other microscopic details.}
\bibitem[{sig()}]{sigma_model_footnote}
\bibinfo{note}{Alternatively, one could use a nonlinear sigma model to describe
  the local-moment degrees of freedom. However, such a description is natural
  only deep inside the ordered phase, where magnitude fluctuations of the
  coarse-grained local-moment variable ${\bm M}$ are small.}
\end{thebibliography}

\end{document}